\documentclass[8pt,a4paper]{article}
\usepackage[english]{babel}
\usepackage[utf8]{inputenc}
\usepackage[cm]{fullpage}
\usepackage[default]{fontsetup}

\usepackage{algorithm}
\usepackage{algpseudocode}
\floatname{algorithm}{Algorithm}
\usepackage{graphicx}
\usepackage{calc}

\title{Fast Computation of Lowest Common Ancestor in Version Control Commit Graph}
\author{
    Stepan Polokhin \\ {\small ilikepugs@yandex-team.ru} \\ {\small Yandex}
    \and Ruslan Savchenko \\ {\small savrus@yandex-team.ru} \\ {\small Yandex}
    \and Yuriy Chernyshov \\ {\small thegeorg@yandex-team.ru} \\ {\small Yandex}
}
\date{}

\newcommand{\sP}{\mathcal{P}}
\newcommand{\sC}{\mathcal{C}}
\newcommand{\sA}{\mathcal{A}}
\newcommand{\sAns}{\mathcal{A}_{ns}}
\newcommand{\sS}{\mathcal{S}}
\newcommand{\slca}{\mathcal{LA}}
\newcommand{\rsq}{\rightsquigarrow}
\newcommand{\rsqns}{\stackrel{ns}\rsq}
\newcommand{\lns}{l_{ns}}

\newcommand{\wG}{\widetilde{G}}
\newcommand{\wB}{\widetilde{B}}
\renewcommand{\ge}{\geqslant}
\renewcommand{\le}{\leqslant}

\newenvironment{salgorithm}[2]{%
\begin{algorithm}
\caption{#1}\label{alg:#2}
\begin{algorithmic}[1]
}{%
\end{algorithmic}
\end{algorithm}
}
\newcommand{\strue}{\textbf{true}}

\newcommand{\snot}{\textbf{not}}
\newcommand{\sand}{\textbf{and}}
\newcommand{\sor}{\textbf{or}}
\newcommand{\sif}{\textbf{if}}
\newcommand{\selse}{\textbf{else}}

\newcommand{\sforall}{\textbf{for all}}
\newcommand{\scontinue}{\textbf{continue}}
\algrenewcommand{\algorithmiccomment}[1]{\hskip3em// #1}

\usepackage{amsthm}
\newtheorem{theorem}{Theorem}
\newtheorem{corollary}[theorem]{Corollary}
\newtheorem{lemma}[theorem]{Lemma}
\newtheorem{fact}[theorem]{Fact}

\newenvironment{slemma}[1]{%
\begin{lemma}\label{lm:#1}
}{%
\end{lemma}
}
\newenvironment{stheorem}[1]{%
\begin{theorem}\label{thm:#1}
}{%
\end{theorem}
}
\newenvironment{scorollary}[1]{%
\begin{corollary}\label{cor:#1}
}{%
\end{corollary}
}
\newenvironment{sfact}[1]{%
\begin{fact}\label{fact:#1}
}{%
\end{fact}
}

\newcommand{\sfigure}[3]{%
  \begin{figure}[h]
  \begin{center}
  \includegraphics{#2}
  \end{center}
  \caption{#3}\label{fg:#1}
  \end{figure}
}
\newcommand{\stwofigure}[4]{%
  \begin{figure}[h]
  \begin{center}
  \includegraphics{#2} \hspace{1cm} \includegraphics{#3}
  \end{center}
  \caption{#4}\label{fg:#1}
  \end{figure}
}

\newcommand{\ralg}[1]{\ref{alg:#1}}
\newcommand{\rthm}[1]{\ref{thm:#1}}
\newcommand{\rlm}[1]{\ref{lm:#1}}
\newcommand{\rcor}[1]{\ref{cor:#1}}
\newcommand{\rfact}[1]{\ref{fact:#1}}
\newcommand{\rfig}[1]{\ref{fg:#1}}
\newcommand{\rsec}[1]{\ref{sec:#1}}
\newcommand{\lsec}[1]{\label{sec:#1}}

\begin{document}
\maketitle

\begin{abstract}
The problem of finding the lowest common ancestor in the commit graph arises when a user wants to merge two (or more) development branches: for example, to add their own development into a common branch.
At that moment, the version control system must determine the merge base~--- the common ancestor of the latest commits of the branches, in order to show the mutual difference between the branches.
The problem is complicated by the fact that the commit graph is not a tree but an arbitrary acyclic graph. Moreover, if there is no common ancestor, searching for one may require traversing the entire graph.
Traversing a graph with several million vertices takes tens of seconds, which becomes noticeable to users.

We show that the commit graph has additional structure~--- supporting branches~--- and that this can be used to speed up the algorithm.
By extracting supporting branches, the graph can be preprocessed and the search for the lowest common ancestor can be significantly accelerated.
Moreover, if removing the supporting branches decomposes the graph into connected components of bounded size, the lowest common ancestor query can be answered in time comparable to searching for a common ancestor within a single component only.

We present an algorithm running in time $O(\delta_Gm_c + n_s\log^2 n_s + m_s\log n_s)$ and requiring $O(d)$ preprocessed data per vertex, where $\delta_G$ is the maximum out-degree of a vertex in $G$, $n_c$ and $m_c$ are the maximum numbers of vertices and edges in a connected component formed after removing the supporting branches, $n_s$ is the number of supporting branches, $m_s$ is the number of edges in the reachability graph between the supporting branches, and $d$ is the maximum number of supporting branches adjacent to any single component.
\end{abstract}


\section{Basic Notions}

Let a directed graph $G=(V,A)$ be given, with vertex set $V$ and edge set $A$, $|V|=n$, $|A|=m$.
A vertex $u$ such that $(u,v)\in A$ is called a \emph{parent} of the vertex $v$.
Accordingly, the vertex $v$ is called a \emph{child} of the vertex~$u$.
We denote the sets of all parents and children of $v$ by $\sP(v)$ and $\sC(v)$, respectively.
Vertices with no parents are called \emph{roots}.
A vertex $u$ is called an \emph{ancestor} of the vertex $v$ if there exists a non-empty path from $u$ to $v$ in $G$.
When speaking of such a path, we write $u\rsq v$.
We denote the set of all ancestors of $v$ by $\sA(v)$.

The \emph{height} of a vertex $v$ in $G$ is the maximum distance from a root to $v$; we denote it by $h(v)$. One can define it recursively as $h(v) = \max(h(\sP(v))) + 1$, $h(\emptyset) = 0$. Here we use the extension of functions to sets that is customary in algorithmic graph theory: $f(X) := \{f(x) \mid x\in X\}$ (from now on we consider all functions extended to sets).
It follows from the definition of $h$ that $h(v) > h(u)$, $\forall u\in\sA(v)$.

A vertex $w\in\sA(u)\cap\sA(v)$ is called a \emph{common ancestor} of the vertices $u$, $v$. From now on we denote $\sA(u,v)=\sA(u)\cap\sA(v)$. A \emph{lowest common ancestor} of $u,v$ is a vertex $w\in\sA(u,v)$ such that $\nexists w'\in\sA(u, v), w\in\sA(w')$.
In other words, $\sA$ induces a partial order on $V$, and we want to find the least vertex of $\sA(u, v)$ with respect to this order. There may be more than one such vertex, and any of them is acceptable as an answer to the problem.
We denote the set of all lowest common ancestors of $u$ and $v$ by $\slca(u,v)$.

Let us state several properties of the commit graph that we will rely on hereafter.

\begin{sfact}{fixedparent}
The list of parents of a vertex is fixed at the moment of its creation and does not change afterwards. New edges are added only to new vertices.
\end{sfact}

Fact~\rfact{fixedparent} allows us to maintain $h(v)$ dynamically: to compute the height, it suffices, when adding a vertex, to traverse its parents and compute $h(v)$ using the formula from the definition.

\begin{sfact}{smallparent}
The number of parents of a vertex in the commit graph is bounded. An ordinary commit has only one parent. Merge commits have two. More than two is extremely rare. We assume $|\sP(v)| = O(1)$, $\forall v\in V$.
\end{sfact}

Fact~\rfact{smallparent} bounds the in-degree of a vertex in the commit graph. The out-degree, however, can be large. We denote the maximum out-degree by $\delta_G$, i.e., $\delta_G = \max_{v\in V}(|\sC(v)|)$.

\begin{sfact}{parentmix}
No parent of a vertex can simultaneously be an ancestor of another parent of that vertex. If a user attempts to make such a merge, the version control system sees that one of the commits being merged is a descendant of the other, and takes only the descendant.
\end{sfact}

\section{A Basic Algorithm for Finding the Lowest Common Ancestor}\lsec{basic}

To find the lowest common ancestor of two vertices $u, v$, one can use Algorithm~\ralg{alg1}. The algorithm traverses~\cite{Schrijver} the vertices in order of decreasing height, starting simultaneously from $u$ and $v$. For the traversal, the algorithm uses a priority queue~\cite{Williams}, with the height serving as the priority.
Initially, the queue contains $u$ and $v$. When a vertex is extracted from the queue, it is marked as visited and its parents are added to the queue.

The algorithm also remembers from which start vertex ($u$ or $v$) it reached the visited vertex.
We denote the vertices visited when starting from $u$ by $visited_u$ (respectively, $visited_v$ when starting from $v$), and those that are only in the queue by $frontier_u$ (respectively, $frontier_v$). We call the set $visited_u \cup frontier_u$ the \emph{traversal ball} from $u$, and $visited_v \cup frontier_v$ the traversal ball from $v$.

The pseudocode of the algorithm is given below. The algorithm is easily generalized to an arbitrary number of start vertices.

\begin{salgorithm}{Finding the lowest common ancestor}{alg1}
\State $PQ \gets (h(u), u, u), (h(v), v, v)$
\State $visited \gets \{\}$
\While{$PQ \ne\emptyset$}
    \State $h, w, start \gets extract\_max(PQ)$
    \State $visited[start, w] \gets \text{\strue}$
    \If {$visited[\overline{start}, w]$} \Comment \text{$\overline{start} \gets v \text{\ \bf if\ } start = u \text{\ \bf else\ } u$}
        \State \Return $w$
    \Else
        \ForAll{$p \in \sP(w)$}
            \State $PQ \gets (h(p), p, start)$
        \EndFor
    \EndIf
\EndWhile
\end{salgorithm}

Let us now prove the correctness of the algorithm and estimate its running time. We introduce an auxiliary notation $\sA_{>h}(v) = \{u\in\sA(v)|h(u) > h\}$, which restricts the set of ancestors to those whose height is greater than $h$. Similarly, we define $\sA_{\ge h}(v) = \{u\in\sA(v)|h(u) \ge h\}$. 

\begin{slemma}{alg1}
At the moment a vertex $w$ is visited, i.e., when $visited[\cdot, w]$ becomes equal to \strue, all vertices from $\sA_{>h(w)}(u) \cup \sA_{>h(w)}(v)$ have already been visited.
\end{slemma}
\begin{proof}
Algorithm~\ralg{alg1} visits vertices in order of non-increasing $h$: at each step, the vertex with the maximum height is extracted from the queue, and its parents, which have smaller heights, are added to the queue. Hence, a vertex with height greater than that of some already processed vertex will never appear in the queue. Moreover, for each vertex $w'\in \sA_{>h(w)}(u) \cup \sA_{>h(w)}(v)$, one can construct a path to the start vertex $u$ or $v$ and show by induction that all vertices on this path, starting from the start vertex, were added to the queue and visited earlier.
\end{proof}

The following follows directly from Lemma~\rlm{alg1}:

\begin{stheorem}{alg1}
Algorithm~\ralg{alg1} finds the lowest common ancestor. The running time of Algorithm~\ralg{alg1} is $O((|\sA_{\ge h}(u)\cup\sP(\sA_{\ge h}(u))| + |\sA_{\ge h}(v)\cup\sP(\sA_{\ge h}(v))|)\log n)$, where $h$ is the height of the found lowest common ancestor.
\end{stheorem}

\sfigure{alg1}{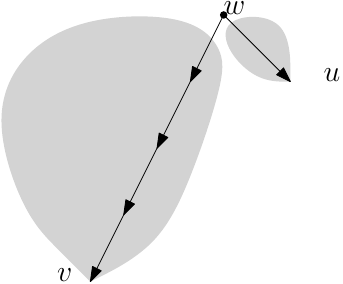}{If $h(v)$ is much larger than $h(u)$, the traversal ball of Algorithm~\ralg{alg1} from $v$ is much larger than the traversal ball from $u$.}

\begin{proof}
If $\sA(u,v)\ne\emptyset$, Algorithm~\ralg{alg1} reaches a vertex from $\sA(u,v)$, since it traverses the sets $\sA(u)$ and $\sA(v)$ entirely until it finds a common ancestor. If the algorithm found a common ancestor $w$, it follows from Lemma~\rlm{alg1} that it is the lowest one. Again by Lemma~\rlm{alg1}, at this moment the algorithm has visited $\sA_{> h}(u)$ and $\sA_{> h}(v)$, spending $O(|\sA_{\ge h}(u)| + |\sA_{\ge h}(v)|)$ iterations on this. In addition to these iterations, the queue still contains unvisited vertices from $\sP(\sA_{\ge h}(u))\cup\sP(\sA_{\ge h}(v))$, whose insertion into the queue must also be accounted for in the running time. The time per operation with the queue is $O(\log n)$ per vertex if a Fibonacci heap~\cite{Fredman} is used as the queue. Putting everything together, we obtain the statement of the theorem.
\end{proof}

Note that Algorithm~\ralg{alg1} requires $O(1)$ preprocessed memory per vertex---to store the height $h$.

Let us state one more consequence of Lemma~\rlm{alg1} that will be useful later:
\begin{scorollary}{path}
Let $w$ be a vertex such that $w\rsq v$ and $h(w)$ is greater than the height of some vertex from $visited_v$. Then $w$ has also been visited by this moment.
\end{scorollary}
\begin{proof}
Let $w'\in visited_v$ be such that $h(w) > h(w')$. By assumption, $w$ is an ancestor of $v$, i.e., $w\in\sA(v)$. Adding the height restriction, we get $w\in\sA_{> h(w')}(v)$. Lemma~\rlm{alg1} states exactly that such a $w$ has already been visited.
\end{proof}

Theorem~\rthm{alg1} implies the main drawback of Algorithm~\ralg{alg1}: if $h(u)$ and $h(v)$ differ significantly, the algorithm needs to traverse a rather large number of vertices; see Fig.~\rfig{alg1}. In practice, this can be many orders of magnitude more than the search for the lowest common ancestor in the typical case where the heights of $u$ and $v$ are close to each other and the common ancestor is located not far from them. In subsequent sections we present a more efficient algorithm. But first we have to introduce additional structure on the graph.

\section{Partitioning the Commit Graph by Supporting Branches}\lsec{break}

The commit graph has structure imposed by development branches: typically, a developer does their local work in a branch forked off some more general branch, and then merges the changes back into it. Thus, the graph contains a certain number of distinguished branches from which other branches fork off and later merge back into them; see Fig.~\rfig{supporting}. We will use this intuition and give a definition of a family of branches that partitions the graph well enough. We will assume that such a family exists in the commit graph.

\sfigure{supporting}{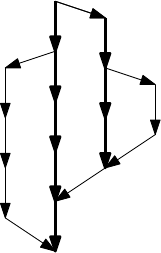}{A commit graph with distinguished branches.}

We call a family of branches (i.e., paths in the graph) $\sS$ in the graph $G$ \emph{supporting} if the following conditions hold:
\begin{enumerate}
    \item Every vertex belongs to at most one branch from $\sS$.
    \item Every root in $G$ belongs to some branch from $\sS$.
    \item After removing all vertices belonging to branches from $\sS$, the graph $G$ decomposes into connected components of bounded size.
    \item For each such component, the set of branches from $\sS$ adjacent to this component (i.e., such that in the original graph there is an edge between the branch and the component) is also bounded.
\end{enumerate}

We will also call the branches of a supporting family, and the vertices on these branches, \emph{supporting}.
Denote $n_s=|\sS|$, $n_c=\max_{C\in CC(G-\sS)} |V_C|$, $m_c=\max_{C\in CC(G-\sS)} |A_C|$, where $CC$ is the function returning the connected components of a graph, and $d=\max_{C\in CC(G-\sS)}|\{\alpha | \alpha\in\sS,\alpha \cap (\delta(C)) \ne\emptyset\}|$, where $\delta(C)=\delta^{in}(C) \cup \delta^{out}(C)$ is the set of vertices adjacent to the subgraph $C$. We assume that for the families of supporting branches we consider, $d\ll n_s\ll n_c\ll n$ holds.

\sfigure{nonsupporting}{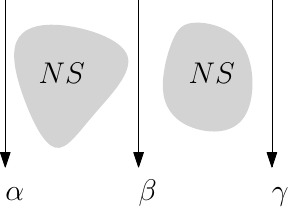}{Supporting branches and non-supporting vertices between them.}

Vertices that are not supporting will be called \emph{non-supporting}. Denote the sets of supporting and non-supporting vertices by $S$ and $NS$, respectively. For a vertex $v\in S$, define $\sigma(v)$ to be the supporting branch to which $v$ belongs. We will use lowercase Greek letters to denote supporting branches.

A path in $G$ all of whose intermediate vertices (i.e., other than the start and the end) are non-supporting is called a \emph{non-supporting} path. We say that $v$ is reachable from $u$ via a non-supporting path if there exists a non-supporting path $u\rsqns v$ in $G$. When speaking of traversing the graph along reversed edges (i.e., from a child to its parent, as in Algorithm~\ralg{alg1}), we will also say that $u$ is reachable from $v$. Denote $m_s=|\{(\alpha, \beta) | \alpha,\beta \in \sS, \exists u,v: u\in\alpha, v\in\beta, u\rsqns v\}|$, i.e., the number of pairs of supporting branches whose vertices are connected by at least one non-supporting path.

Denote by $\sA_{ns}(v)$ the set of all non-supporting $u\in NS$ such that $v$ is reachable from $u$ via a non-supporting path, i.e., $\sA_{ns}(v) = \{u|u\in NS, u\rsqns v\}$. Similarly, for a supporting branch $\alpha\in\sS$, denote by $\sA_\alpha(v)$ the set of all $u\in\alpha$ from which $v$ is reachable via a non-supporting path: $\sA_{\alpha}(v) = \{u|u\in\alpha, u\rsqns v\}$.

We have finished introducing the family of supporting branches. All further algorithms in this paper will rely on the existence of such a family. The main idea of why it is useful to us: the supporting branches partition the graph into non-supporting connected components, each bordering a small number of supporting branches; see Fig.~\rfig{nonsupporting}.

\section{Heuristic Speedup of Finding the Lowest Common Ancestor Using Supporting Branches}\lsec{pruning}

Let us make use of the fact that we have supporting branches to speed up the graph traversal when finding the lowest common ancestor.
We will try to direct the search along the supporting branches and not leave them unless there is an explicit need to do so (see Fig.~\rfig{pruned}).
Before we can state the condition that will allow us not to add a vertex to the traversal queue, we will need to extend the preprocessed information we store.

\sfigure{pruned}{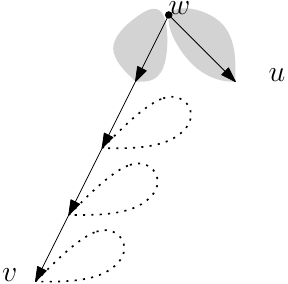}{Intuition behind the traversal speedup: ignore vertices that will certainly not help reach the goal.}

In addition to $h(v)$, for each vertex we will store:
\begin{enumerate}
  \item $\lns(v) = min(h(\sA_{ns}(v)))$---the minimum height that the search can reach by visiting only non-supporting vertices.
  \item $m_\alpha(v) = max(h(\sA_\alpha(v)))$---the maximum height on the supporting branch $\alpha$ that the search can reach by visiting only non-supporting vertices.
\end{enumerate}

\begin{slemma}{mem1}
Storing $\lns$ and $m_\alpha$ requires $O(d)$ memory per vertex.
\end{slemma}
\begin{proof}
Storing $\lns$ adds $O(1)$ memory for each vertex, so we only need to deal with $m_\alpha$.
Note that $m_\alpha$ can be stored only for non-supporting vertices, while for supporting ones it can be computed using $O(1)$ non-supporting parents. Since the number of branches reachable (via reverse edges) from a non-supporting vertex is at most $d$, each non-supporting vertex requires $O(d)$ memory to store $m_\alpha$.
\end{proof}

It is easy to see that computing $\lns(v)$ and $m_\alpha(v)$ only requires knowing the values of these functions in the parents at the moment the vertex is added to the graph. That is, maintaining these functions requires $O(d)$ time per added vertex.

Let us prove two structural lemmas that will help us prune unpromising candidates for the traversal.

\begin{slemma}{prunel}
Let $v'$ be a vertex that the traversal reached from the starting vertex $v$, and let $frontier_u$ be the current set of vertices in the queue $PQ$ from the traversal ball from $u$. Then, if $\lns(v') > max(h(frontier_u))$, the set $\sA_{ns}(v')$ does not contain the lowest common ancestor of $u$ and $v$.
\end{slemma}
\begin{proof}
By Lemma~\rlm{alg1}, the height of the vertices visited by the traversal is non-decreasing, and therefore the height of the lowest common ancestor is certainly not greater than $max(h(frontier_u))$. By the assumption of the lemma, $max(h(frontier_u)) < \lns(v')$, and $\lns(v') = min(h(\sA_{ns}(v')))$ by definition. Consequently, $\sA_{ns}(v')$ does not contain the lowest common ancestor of $u$ and $v$.
\end{proof}

\sfigure{prunel}{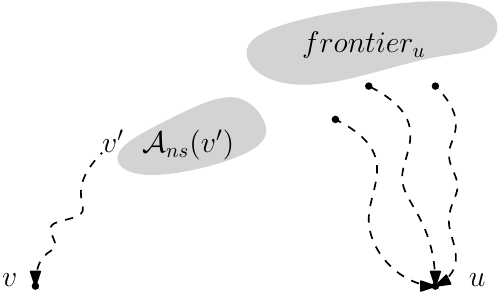}{If the set of vertices that we are going to visit in one branch of the traversal has height greater than any vertex from the frontier of the other branch of the traversal, then the search will certainly not converge within this set.}

\begin{slemma}{prunem}
Let $v'$ be a non-supporting vertex that is a parent of some vertex $x$ on a supporting branch, which the traversal reached from the starting vertex $v$, and let $m_\alpha(v') \le m_\alpha(v'')$ for some other vertex already visited within the traversal from $v$. Let $w$ be the lowest common ancestor of $u$ and $v$. Then, if the path $w\rsq v$ passes through a vertex on the supporting branch $\alpha$, skipping $v'$ when considering the parents of $x$ will not affect the reachability of $w$ by the traversal ball from $v$.
\end{slemma}
\begin{proof}
Let $w'\in\alpha$ be such that $h(w') = m_\alpha(v'')$. Then, if $v' \notin w'\rsqns v''$, removing $v'$ from consideration will not worsen the search space: by the assumption, we know that via $v'$ we can reach either $w'$ or an ancestor of $w'$ on $\alpha$, but by no means a descendant, and therefore the set of reachable vertices on $\alpha$ will not deteriorate.
If, however, $v' \in w'\rsqns v''$, then there exists a path $v'\rsqns v''$, and therefore the traversal will reach $v'$ bypassing $x$, so skipping $v'$ when considering the parents of $x$ will not affect anything.
\end{proof}

\stwofigure{prunem}{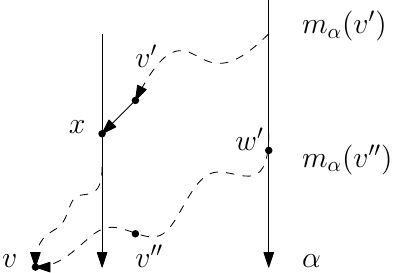}{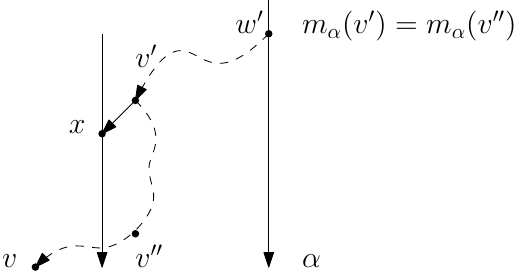}{If, when passing through a vertex, the candidates from the supporting branch $\alpha$ do not improve, then this vertex can be skipped from the viewpoint of reaching $\alpha$.}

Now we can state an algorithm that will avoid visiting unpromising vertices. Algorithm~\ralg{alg2} is a modification of Algorithm~\ralg{alg1} which, when considering the list of parents of a supporting vertex, skips (i.e., does not put into the queue) non-supporting vertices $v'$ from the traversal ball for $v$ (respectively, $u'$ from the traversal ball for $u$) satisfying both of the following conditions:
\begin{enumerate}
  \item $\lns(v') > max(h(frontier_u))$ (respectively, $\lns(u') > max(h(frontier_v))$).
  \item $m_\sigma(v') \le max(m_\sigma(visited_v)) \forall\sigma\in\sS$ (respectively, $m_\sigma(u') \le max(m_\sigma(visited_u)) \forall\sigma\in\sS$).
\end{enumerate}

\begin{salgorithm}{Finding the Lowest Common Ancestor with Pruning of Non-Supporting Candidates}{alg2}
\Function{Prune}{$p, start$}
    \State\Return $\lns(p) > max(h(frontier_{\overline{start}}))$ \text{\sand} $m_\sigma(p) \le max(m_\sigma(visited_{start})) \text{ \sforall\ } \sigma\in\sS$
\EndFunction
\State
    \State $PQ \gets (h(u), u, u), (h(v), v, v)$
    \State $visited \gets \{\}$
    \While{$PQ \ne\emptyset$}
        \State $h, w, start \gets extract\_max(PQ)$
        \State $visited[start, w] \gets \text{\strue}$
        \If {$visited[\overline{start}, w]$}
            \State \Return $w$
        \Else
            \ForAll{$p \in \sP(w)$}
                \If {\text{\snot} \textsc{Prune}(p, start) \text{\sor} $\sigma(p) = \sigma(w)$}
                    \State $PQ \gets (h(p), p, start)$
                \EndIf
            \EndFor
        \EndIf
    \EndWhile
\end{salgorithm}

\begin{stheorem}{alg2}
If $\sA(u,v)\ne\emptyset$, then Algorithm~\ralg{alg2} will find the lowest common ancestor of $u$ and $v$.
\end{stheorem}
\begin{proof}
Algorithm~\ralg{alg2} differs from Algorithm~\ralg{alg1} by the addition of pruning of unpromising vertices. We need to show that, from the viewpoint of the traversal, Algorithm~\ralg{alg2} will visit all the vertices that lead Algorithm~\ralg{alg1} to finding the lowest common ancestor of the vertices $u$ and $v$.

Suppose that at some step, when visiting a vertex from the traversal ball for $v$, we did not add the vertex $v'$ to the queue.
This means that both conditions from $\textsc{Prune}$ were satisfied.
By Lemma~\rlm{prunel}, the desired lowest common ancestor certainly does not lie in $\sA_{ns}(v')$, and skipping this set will not affect the result. Also, by Lemma~\rlm{prunem}, skipping $v'$ will not affect the reachability of the other supporting branches: the traversal will reach them, and moreover it will visit vertices with height certainly not smaller than the one it would have reached by passing through $v'$, or it will pass through $v'$ in a different way.
\end{proof}


\subsection{Skipping Unpromising Candidates on a Supporting Branch}

Algorithm~\ralg{alg2} can be more efficient than Algorithm~\ralg{alg1} in practice, since it is able to exclude non-supporting vertices from the traversal, but it still has a drawback: it does not skip vertices on the supporting branches, and there may be quite many of them. We would like to jump over supporting vertices that will certainly not help in finding the lowest common ancestor.

We already have the function $\textsc{Prune}$, which can be interpreted as follows. For a non-supporting vertex, $\textsc{Prune}$ checks that the traversal ball cannot reach, through it, any vertices of interest to us, i.e., a non-supporting common ancestor or simply supporting vertices that the traversal cannot reach through the already visited vertices. For a supporting vertex, $\textsc{Prune}$ checks that all non-supporting parents can be skipped, and one should jump directly to the parent on the supporting branch.

\sfigure{pruneh}{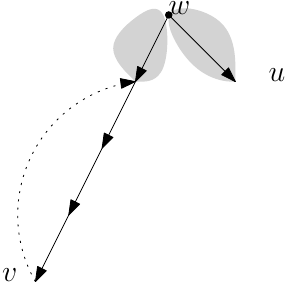}{If the vertices on a supporting branch have no non-supporting parents that are interesting to consider, these vertices can be skipped.}

It follows that if there is a sequence of vertices on a supporting branch with the conditions of $\textsc{Prune}$ satisfied, all of them can be skipped. That is, when considering a vertex on the supporting branch $\alpha$, either the conditions of $\textsc{Prune}$ are violated for it, and we add it to the traversal, or we need to find its nearest ancestor (i.e., the one with the maximum height~$h$) on the supporting branch $\alpha$ for which the conditions of $\textsc{Prune}$ are violated.

Note that the conditions of $\textsc{Prune}$ depend only on $\lns(v')$, $m_\sigma(v')$, $max(h(frontier_u))$, and $max(m_\sigma(visited_v))$ $\forall\sigma\in\sS$. Moreover,
\begin{itemize}
     \item $\lns(v')$ and $m_\sigma(v')$ depend only on the graph, and they can be computed in advance for all supporting vertices (i.e., they do not depend on the current step of the algorithm);
    \item $max(h(frontier_u))$ and $max(m_\sigma(visited_v))$ depend only on the current step of the traversal and do not depend on how far we want to jump along the supporting branch $\alpha$ (given that the conditions of $\textsc{Prune}$ are satisfied for the vertices being skipped).
\end{itemize}

Any data structure that allows finding the minimum or maximum on any prefix will suit us.
We can take a segment tree~\cite{Berg}, which is a binary tree over intervals, in whose nodes the minimum or maximum of the subtree, respectively, is stored. A segment tree allows adding elements to an array and also finding the minimum or maximum for any interval in $O(\log l)$, where $l$ is the length of the original array.

Let $l$ be the maximum length of a supporting branch (recall that we earlier denoted the total number of branches by $s$). The segment tree built for the supporting branch $\alpha$ and $\lns$ allows us to find, in $O(\log l)$, the nearest vertex $v'\in\alpha$ for which $\lns(v') \le max(h(frontier_u))$. Similarly, the segment tree for $\alpha$ and $m_\sigma$ will allow us to find the nearest vertex $v'\in\alpha$ for which $m_\sigma(v') > max(m_\sigma(visited_v))$. By iterating over all $\sigma\in\sS-\alpha$, we obtain the $v'$ to which we can jump without losing search quality.

\begin{stheorem}{pruned}
Algorithm~\ralg{alg2} with skipping along the supporting branches finds the lowest common ancestor in time $O((visited_S + frontier_S) s\log l + (visited_{NS} + frontier_{NS})s + (visited + frontier) \log n) = O(n(s\log l + \log n))$.
\end{stheorem}
\begin{proof}
The correctness of the new version of the algorithm is proved similarly to Theorem~\rthm{alg2}. For each visit of a supporting vertex, we need to spend $O(s\log l)$ to jump to it using $s$ segment trees. For each non-supporting vertex reached by the search, we need to spend $O(s)$ to check the conditions of $\textsc{Prune}$. Additionally, all operations with the queue $PQ$ must be taken into account.
\end{proof}

In a naive implementation, each tree is built for all vertices of the supporting branch, which requires $O(s)$ additional memory per vertex.
In Section~\rsec{memory}, we will show how the memory consumption can be reduced from $O(s)$ to $O(d)$ per vertex. For now, let us examine how to reduce the factor from $s$ to $d$ in the expression for the algorithmic complexity of adding a new vertex to the graph.

\begin{stheorem}{dtime}
Adding a new vertex to the graph requires $O(d\log l)$ time to maintain the tree structure.
\end{stheorem}
\begin{proof}
Consider adding a new supporting vertex (the trees are built only for supporting vertices). By Fact~\rfact{smallparent}, the vertex has $O(1)$ non-supporting parents. By the construction of the supporting family, each of them is reachable from only $O(d)$ supporting branches. It follows that only $O(d)$ coefficients $m_\sigma$, $\sigma\in\sS$, are defined for the new vertex, and the trees need to be updated only for them.
\end{proof}



\section{An Algorithm with Guaranteed Running Time}\label{sec:guaranteed}

Algorithm~\ralg{alg2} with pruning is heuristic: it works excellently in practice, but has no good theoretical guarantees on the running time and can potentially run for a long time. In this section, we present an algorithm that runs in time $O(\delta_Gm_cd + n_s\log n_s + m_s)$ and requires $O(n_s)$ memory per vertex (the constants used in the formula were defined earlier in Section~\rsec{break} for the partition by supporting branches). Later, in Section~\rsec{memory}, we will show how to reduce the memory consumption to $O(d)$ per vertex, at a slight cost in running time (a logarithmic factor will be added).

First, construct an auxiliary graph for $G$ for the chosen family of supporting branches $\sS$. We denote the resulting graph by $\wG$. The vertex set $V\wG$ coincides with $VG$. Before constructing the edge set $A\wG$, we introduce an additional notation.

For a supporting vertex $v$ on a supporting branch $\alpha$ and for a supporting branch $\beta$, define the maximum height $m^*_\beta(v)$ of a vertex on the supporting branch $\beta$ from which either $v$ or some ancestor of $v$ on $\alpha$ is reachable via non-supporting paths. Formally, $m^*_\beta(v) = \max(h(\cup_{v'\in\sA^*(v)\cup\{v\}} \sA_\beta(v')))$, where $\sA^*(v) := \sA(v)\cap\sigma(v) = \sA(v)\cap\alpha$. That is, $m^*_\beta(v)$ is the maximum height that the traversal ball can reach on $\beta$ when passing through $v$ only along the supporting branch $\alpha$ and through non-supporting vertices. If $m^*_\beta(v)$ is attained at a vertex from $\sA_\beta(v)$, we say that $\beta$ is \emph{maximally reachable} from $v$.

Now we construct the edge set $A\wG$. For each vertex $v$ on a supporting branch $\alpha$ and for each other supporting branch $\beta\in\sS-\alpha$, we add an edge:

\begin{enumerate}
  \item $u\rightarrow v, u\in\beta$, if $u\in\sA_\beta(v)$ and $m^*_\beta(v) = h(u)$ (i.e., $\beta$ is maximally reachable from $v$, and $u$ is the corresponding vertex on $\beta$);
  \item $w\rightarrow v, w\in\alpha$, if $\beta$ is not maximally reachable from $v$, and $w$ is the first ancestor of $v$ on $\alpha$ from which $\beta$ is maximally reachable;
  \item $w\rightarrow v, w\in\alpha$, if $w$ is the parent of $v$ on $\alpha$.
\end{enumerate}

An example of constructing $A\wG$ is shown in Fig.~\rfig{tildeg}. It is worth noting that both the edge $(u,w)$ and $(u,v)$ are added, since the maximal reachability condition holds for each of them. Later we will use the fact that all such edges are present from the vertex $u$ in the augmented graph.

\sfigure{tildeg}{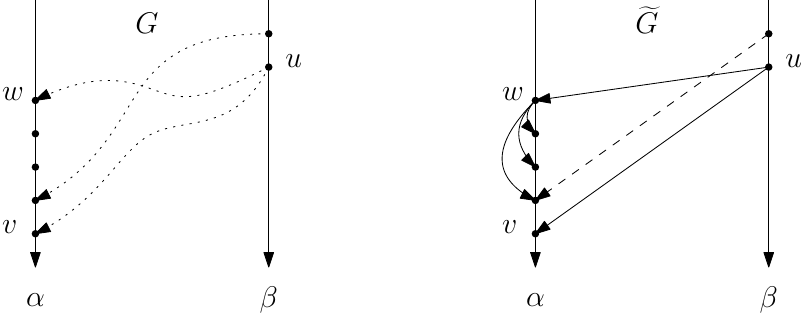}{Graph augmentation: on the left, the graph $G$; on the right, the augmentation $\wG$. The added edges are shown as solid lines; the dashed edge is not added (because it does not satisfy the maximal reachability condition).}

In words, we add the transition $u\rightarrow v$ between supporting branches in all cases where it would bring the traversal ball from $v$ to $u$ on the supporting branch $\sigma(u)$ at a height not less than the one at which the traversal could reach $\sigma(u)$ through some ancestor of $v$ on $\sigma(v)$.
We call such augmentations augmentations of the first kind.
In addition to them, along each supporting branch, transitions to the nearest ancestor with an augmentation of the first kind are added---these will be augmentations of the second kind.
We also keep the edges from the parents on the same supporting branch in order to preserve the connectivity of the supporting branch in the augmented graph. These are augmentations of the third kind.
The union of all these augmentations yields the graph $\wG$.

It is easy to see that in the graph $\wG$, for a vertex $v\in\alpha$, if for some ancestor $w$ of $v$ on $\alpha$ there is an augmentation of the first kind from the supporting branch $\beta$, then $\beta$ can be reached from $v$ in at most two transitions.
It may seem that augmentations of the second kind should have been made to directly connect different supporting branches. Indeed, from the standpoint of constructing the subsequent algorithm, that would be simpler, but the chosen approach will later be optimized for memory. Namely, in Section~\rsec{memory} we will merge augmentations of the second kind constructed for different supporting branches. For now, for clarity of presentation, we will work with $\wG$ as is, without memory optimizations. If the number of supporting branches $n_s$ is small and the memory optimization is not needed, one could indeed make the augmentations of the second kind directly between supporting branches.

In order to hide, in our reasoning and in the pseudocode we present, the need to think about whether one or two transitions are required to get to another supporting branch, we introduce the notation $\wB(v)$ for the set of all $u$ such that in $\wG$ there is an augmentation of the first kind $(u, v)$ or an augmentation of the first kind $(u, w)$ together with an augmentation of the second kind $(w, v)$. In other words, $\wB$ is the set of vertices on other supporting branches with maximum height from which $v$ is reachable. Formally, $\wB(v) = \{u| u\in S, h(u) = m^*_{\sigma(u)}(v)\}$.

Note that $\wG$ and $\wB$ are not hard to maintain dynamically. Relying on Fact~\rfact{fixedparent}, we show that $m^*_\beta(v)$ for $v\in\alpha$ can be computed from the parents of $v$ alone. Indeed, for this it suffices to know $m_\beta(v)$, which is computed from the parents, and the common maximum over the branch $\alpha$, which can be stored in the supporting parent of $v$. It remains only to store, together with $m^*_\beta(v)$, the corresponding $u\in\beta$---this way we construct both $\wG$ and $\wB$.

Let us prove a useful lemma that justifies using the height of vertices when traversing the graph $\wG$.
The height in the graph $\wG$ may differ by more than one, but our traversal algorithms never use this anywhere. Only the comparison of heights with each other matters.
\begin{slemma}{wheight}
The height $h$ in the graph $G$ is a valid height also for the graph $\wG$: for any vertex $u\in \wB(v)$, $h(v) > h(u)$ holds.
\end{slemma}
\begin{proof}
By the construction of $\wG$ and the definition of $\wB(v)$, there exists a path $u\rsq v$ in $G$. Hence, $u\in\sA(v)$, and therefore $h(v) > h(u)$.
\end{proof}

\medskip
We have constructed the augmented graph $\wG$. Recall that we need it to find the lowest common ancestor of the vertices $u$ and $v$ in the graph $G$. The algorithm that we consider in this section consists of several steps:
\begin{itemize}
  \setlength{\itemindent}{\widthof{\textbf{Step 4.}}}
  \item[\textbf{Step 1.}] Construct $S(u) = \{u'\in\sA_{ns}(u) \cap S | h(u') = m_{\sigma(u')}(u)\}$. In other words, $S(u)$ is the set of supporting vertices of maximum height (i.e., one per supporting branch) from which $u$ is reachable via non-supporting paths. Similarly, construct $S(v)$.
  \item[\textbf{Step 2.}] Find the lowest common ancestor of the vertices from $S(u)$ and $S(v)$ in the augmented graph $\wG$. The search stops when the traversal balls $visited_u$ and $visited_v$ intersect on some supporting branch.
  \item[\textbf{Step 3.}] If a common ancestor was found at step 2, collect, among its descendants, candidates for the lowest common ancestor of $u$ and $v$.
  \item[\textbf{Step 4.}] Check whether any of the candidates found at the previous step is the lowest common ancestor.
\end{itemize}

\sfigure{notwcandidate}{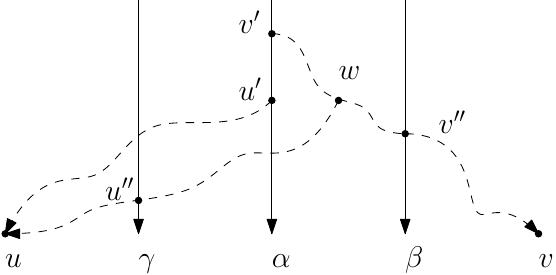}{A supporting branch $\alpha$ on which the search balls from $u$ and $v$ met, and their lowest common ancestor $w$, which does not lie on $\alpha$.}

The intuition for our algorithm can be seen in Fig.~\rfig{notwcandidate}: if the vertices $u$ and $v$ have common ancestors, then there is a common ancestor on a supporting branch (we prove this carefully below). But a common ancestor on a supporting branch is not necessarily the lowest common ancestor, and it must be looked for among the non-supporting descendants of the common ancestor already found. Moreover, the lowest common ancestor is not necessarily reachable from those supporting branches by which the search balls first reached the common supporting branch, so we need to approach the search for possible candidates carefully.

The algorithm begins with constructing $S(u)$ and $S(v)$. To construct $S(v)$, one can traverse $\sA_{ns}(v)$ in order of non-increasing height, starting from $v$: the first vertex on a supporting branch reached by the traversal will lie in $S(v)$. In practice, it is useful to combine this traversal with Algorithm~\ralg{alg1}---then, if $u$ and $v$ turn out to be in the same non-supporting connected component, the lowest common ancestor can be found immediately. Alternatively, one can store $S(v)$ precomputed for each vertex $v$---this requires $O(d)$ memory per vertex. To compute $S(v)$ when a new vertex is added, it suffices to know $S(\cdot)$ and $m_\sigma(\cdot)$ for its parents. This completes step~1.

In the following subsections we consider steps~2--4 in detail.


\subsection{Finding the Lowest Common Ancestor in the Augmented Graph (Step~2)}

At this step, we have the start supporting vertices from $S(u)\subseteq\sA_{ns}(u)$ and $S(v)\subseteq\sA_{ns}(v)$, and on each supporting branch we have taken the vertex with the maximum height.
We want to find a common ancestor of the vertices $u$ and $v$ among the supporting vertices of maximum height (i.e., the lowest supporting common ancestor). For this, we use a modification of Algorithm~\ralg{alg1} with the following differences:
\begin{enumerate}
  \item The search starts not from a single point, but from several. For this, we add them to the queue at initialization, with the height as the priority. They will be extracted from the queue in priority order, possibly yielding their place to some new vertices.
  \item The task of the traversal at this step is to traverse the supporting branches. Therefore, we will add only vertices on not-yet-visited supporting branches for each traversal ball.
  \item The algorithm stops when a common supporting branch is found in both traversal balls. In this case, the vertices on this branch reached by the traversal may be different.
\end{enumerate}

The pseudocode of the algorithm is presented below. The algorithm returns the found supporting branch and a hash table $reach$, in which, for each traversal ball and each reached branch, the vertex at which the reach occurred is stored.

\begin{salgorithm}{Step 2: finding reachable supporting branches in the graph $\wG$}{step2}
\State $PQ \gets (h(u'), u', u) \text{\ \sforall\ } u'\in S(u)$
\State $PQ \gets (h(v'), v', v) \text{\ \sforall\ } v'\in S(v)$
\State $reach \gets \{\}$
\While{$PQ \ne\emptyset$}
    \State $h, w, start \gets extract\_max(PQ)$
    \If {$reach[start, \sigma(w)]$}
        \State\scontinue
    \EndIf
    \State $reach[start, \sigma(w)] \gets w$
    \If {$reach[\overline{start}, \sigma(w)]$}
        \State\Return $\sigma(w), reach$
    \EndIf
    \ForAll{$w'\in\wB(w)$}
        \State $PQ \gets (h(w'), w', start)$
    \EndFor
\EndWhile
\end{salgorithm}

\sfigure{wg}{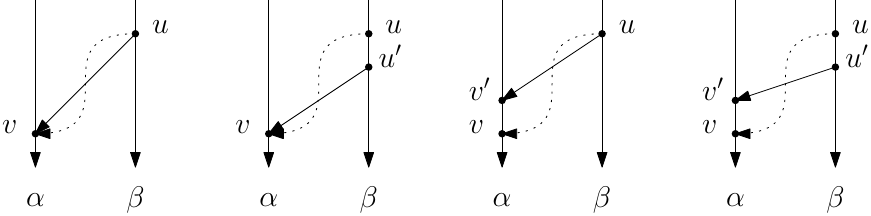}{If supporting vertices are reachable via a non-supporting path in $G$, then they are also reachable in $\wG$.}

Let us prove several statements needed to prove the correctness of the algorithm. First, we prove auxiliary lemmas about $\wG$.

\begin{slemma}{wg}
If for supporting vertices $v\in\alpha$, $u\in\beta$ there exists a non-supporting path $u\rsqns v$ in the graph $G$, then in the graph $\wG$ there exists either an edge $(u,v)$ or an edge $(u', v')$ (possibly $u'=u$ or $v'=v$), where $v'\in\alpha$ and $h(v') \le h(v)$, and $u'\in\beta$ and $h(u') \ge h(u)$.
\end{slemma}
\begin{proof}
By the construction of $\wG$, either there is an edge $(u,v)$, or an edge from a descendant of $u$ on $\beta$ to $v$, or for some ancestor $v'$ of the vertex $v$ on $\alpha$ there is already an edge $(u, v')$ or $(u', v')$, for $u'$ a descendant of $u$ on $\beta$ (see Fig.~\rfig{wg}).
\end{proof}

\sfigure{wgreach}{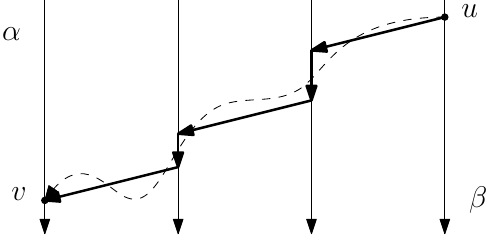}{If in $G$ there is a path between supporting vertices, then in $\wG$ there is also a path between these vertices.}

\begin{slemma}{wgreach}
If in the graph $G$ between supporting vertices $u$ and $v$ there exists a path $u\rsq v$, then in $\wG$ between $u$ and $v$ there also exists a path.
\end{slemma}
\begin{proof}
Consider the path $u\rsq v$ and its intersections with supporting branches (see Fig.~\rfig{wgreach}). By Lemma~\rlm{wg}, each non-supporting segment of the path $u\rsq v$ between two supporting branches can be replaced by a path in the graph $\wG$ (recall that in $\wG$ all edges on supporting branches are preserved---augmentations of the third kind).
\end{proof}

Now we proceed directly to Algorithm~\ralg{step2}. First, we prove for it an analogue of Corollary~\rcor{path}.

\begin{slemma}{step2reach}
If at some moment the traversal ball from $v$ visits a supporting branch $\alpha$ at a vertex with height $h$, then for any supporting branch $\beta$ for which there exists $w\in\beta$ such that $w\rsq v$ in $G$ and $h(w) > h$, one can assert that the branch $\beta$ has already been visited by the traversal ball, and $reach[v,\beta]$ already contains the vertex with the maximum reachable height.
\end{slemma}
\begin{proof}
The proof is analogous to Corollary~\rcor{path}. By Lemma~\rlm{wheight}, Algorithm~\ralg{step2} traverses vertices in order of non-increasing height. By Lemma~\rlm{wgreach}, in $\wG$ there are transitions by which the traversal can reach the required vertex on each such supporting branch. Combining these two facts, we obtain that all such vertices have already been visited.
\end{proof}

\begin{slemma}{step2exists}
If Algorithm~\ralg{step2} did not find a common supporting branch, then no common ancestor exists.
\end{slemma}
\begin{proof}
First, let us show that if $u$ and $v$ have a common ancestor $w$, then there is also a common ancestor on a supporting branch. Recall that we required of the family $\sS$ that every root in $G$ belongs to some supporting branch. Hence, when traversing $\sA(w)$, we will necessarily reach a supporting branch and obtain a supporting common ancestor of $u$ and $v$.

Consider a supporting common ancestor $w$ of the vertices $u$ and $v$. By Lemma~\rlm{wgreach}, in $\wG$ there exist paths $w\rsq u$ and $w\rsq v$. But then, by Lemma~\rlm{step2reach}, the algorithm must visit $\sigma(w)$ with both traversal balls, or stop earlier. In any case, it must find a common supporting branch.
\end{proof}

\begin{slemma}{step2unique}
Let Algorithm~\ralg{step2} have found a common supporting branch. Let $u'$ and $v'$ be the vertices in which the traversal balls reached this supporting branch. Assume, without loss of generality, that $h(v') \le h(u')$. Then either $v'$ is the lowest common ancestor, or one is found among the non-supporting vertices.
\end{slemma}
\begin{proof}
By the assumption of the lemma, $v'\in \sA(u, v)$. It suffices for us to prove that there is no other supporting vertex $w$ such that $w\in\sA(u, v)$ and $v'\in \sA(w)$. But in that case we would have $h(w) > h(v')$, and hence, by Lemma~\rlm{step2reach}, Algorithm~\ralg{step2} should have reached $\sigma(w)$ with both traversal balls earlier, a contradiction.
\end{proof}

\begin{stheorem}{step2}
Algorithm~\ralg{step2} either finds a supporting vertex such that the lowest common ancestor of the vertices $u$ and $v$ is this vertex itself or its non-supporting descendant, or guarantees that there are no common ancestors. The running time of the algorithm is $O(m_s + n_s \log n_s)$.
\end{stheorem}
\begin{proof}
The first part of the statement follows from Lemmas~\rlm{step2exists} and~\rlm{step2unique}. It remains to deal with the running time.

Since the algorithm visits a new supporting branch at each step, there are $O(n_s)$ steps in total. At each step, one needs to maintain a priority queue of $n_s$ elements (formally speaking, there are $m_s$ of them, but for each supporting branch one can maintain in the heap a single vertex with the maximum height), which requires time $O(\log n_s)$ (again using a Fibonacci heap). Additionally, the algorithm needs to consider all possible transitions $\wB(w)$ in the visited vertices $w$, which, summed over all visited vertices, gives $O(m_s)$. Adding everything together, we obtain $O(m_s + n_s \log n_s)$.
\end{proof}

Theorem~\rthm{step2} guarantees that the lowest common ancestor of $u$ and $v$ is either a supporting vertex $v'$ found by Algorithm~\ralg{step2} or some non-supporting descendant of $v'$, and if such a $v'$ was not found, then no common ancestor exists. This completes step~2. After performing steps~3 and~4, we either find such a non-supporting descendant, or make sure that there are none, in which case $v'$ is the actual lowest common ancestor.

\subsection{Finding Candidates for the Lowest Common Ancestor among Non-Supporting Vertices (Step~3)}

We now proceed to step~3. At the previous step, we obtained a supporting branch $\alpha$ and vertices $u',v'\in\alpha$ such that $u'$ lies in the traversal ball of $u$, and $v'$ in the traversal ball of $v$. Without loss of generality, we assume $h(v') \le h(u')$.
At the current step of the algorithm, we need to collect all possible candidates among which the lowest common ancestor is guaranteed to be found (if one exists among the descendants of $v'$). By Theorem~\rthm{step2}, we need to consider only non-supporting vertices.

\sfigure{lcavsbranch}{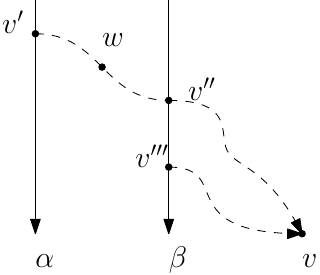}{The position of the lowest common ancestor relative to the reachable supporting branch.}

Let $w$ be the desired lowest common ancestor (see Fig.~\rfig{lcavsbranch}). Then either $w \rsqns v$, or a supporting branch is encountered on the path $w \rsq v$. Let $\beta$ be the first such supporting branch. Let $v''$ be the first vertex on the path $w \rsq v$ from $\beta$; then $w \rsqns v''$ and $v'' \rsq v$. By Lemma~\rlm{step2reach}, the traversal ball reaches $\beta$ either at $v''$ or at $v'''$ with $h(v''') > h(v'')$. That is, $reach[v, \beta]$ will contain a vertex with height at least $h(v'')$.
It follows that $w$ must be sought either among the non-supporting paths between $v'$ and $v$, or among the non-supporting paths between $v'$ and $v''$, where $v''$ must be taken on the reached supporting branch.

\sfigure{components}{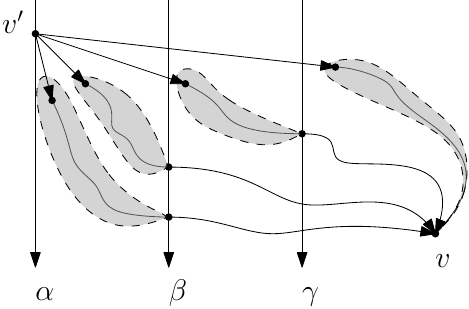}{Several connected components adjacent to the same vertex $v'$, at which the search converges---candidates for being the lowest common ancestor.}

To understand which $v''$ should be taken, consider Fig.~\rfig{components}. It depicts four connected components: two adjacent to the supporting branch $\beta$, a third adjacent to the supporting branch $\gamma$, and a fourth adjacent to the start vertex $v$. The lowest common ancestor may lie in any of them, or in none. Thus, in order to find the lowest common ancestor, we need to consider all such connected components: in the worst case, the lowest common ancestor is in none of them, and then we can conclude that $v'$ is the desired answer (i.e., $v'\in \slca(u, v)$). If other common ancestors exist in these components, we need to find all of them and choose the one with the maximum height.

Combining the reasoning above, we obtain the lemma:
\begin{slemma}{candiddates}
The lowest common ancestor lies either on the path $v'\rsqns v$, or on the path to the supporting vertex $v'\rsqns v''$, where $v''$ lies on a reachable supporting branch and $h(reach[v, \sigma(v'')]) \ge h(v'')$.
\end{slemma}

Thus, to find all candidates for the lowest common ancestor, we need to find all non-supporting vertices $w$ such that, on the one hand, they are reachable from $v'$ via non-supporting paths, and on the other hand, either $v$ or supporting vertices from $\sA(v)$ are reachable from them via non-supporting paths. This can be done with Algorithm~\ralg{step3}.

\begin{salgorithm}{Step 3: finding the set $C$ of candidates among which common ancestors may be found}{step3}
\State $Q \gets \sC(v')\cup NS$
\State $visited_{down} \gets \{x:\text{\strue,\  \sforall\ } x\in Q\}$
\State $R\gets []$
\While{$Q \ne\emptyset$}
    \State $x \gets pop(Q)$
    \ForAll{$y\in \sC(x)\text{ \sand\ \snot\ } visited_{down}[y]$}
        \State $visited_{down}[y] \gets \text{\strue}$
        \If{$y\in NS$}
            \State $Q \gets y$
        \EndIf
        \If{$(y\in S \text{\ \sand\ }h(y) \le reach[v,\sigma(y)])\text{\ \sor\ } y=v$}
            \State $R \gets y$
        \EndIf
    \EndFor
\EndWhile
\State $Q \gets R$
\State $visited_{up} \gets \{x:\text{\strue,\ \sforall\ } x\in Q\}$
\State $C\gets [] \text{\ \sif\ } v \notin R \text{\ \selse\ } [v]$
\While{$Q \ne\emptyset$}
    \State $x \gets pop(Q)$
    \ForAll{$y\in \sP(x)\text{ \sand\ \snot\ } visited_{up}(y)$}
        \State $visited_{up}[y] \gets \text{\strue}$
        \If{$y\in NS\text{ \sand\ } visited_{down}[y]$}
            \State $Q\gets y$
            \State $C\gets y$
        \EndIf
    \EndFor
\EndWhile
\State\Return $C$
\end{salgorithm}

Algorithm~\ralg{step3} consists of the following stages:
\begin{enumerate}
  \item Traverse the non-supporting vertices along edges from parent to child, starting from the non-supporting vertices of $\sC(v')$. Such a traversal proceeds in order of non-decreasing height, whereas all our previous algorithms traversed in order of non-increasing height, along edges from child to parent. This way, we find all vertices reachable from $v'$ via non-supporting paths.
  \item Collect all reachable vertices on supporting branches satisfying the condition: a vertex $x$ on branch $\beta$ is taken only if $\beta$ was reached earlier by Algorithm~\ralg{step2} from $v$, and $h(x) \le h(reach[v,\beta])$. Denote this set by $R$. If the traversal also reached $v$, then we add $v$ to $R$ as well. This way, we obtain the set of all $v''$ that need to be considered.
  \item Traverse the non-supporting vertices along edges from child to parent, starting from $R$, visiting only the vertices already visited at the first stage---the others are of no interest to us. This way, we find all vertices $w$ that are themselves reachable from $v'$ and from which the suitable vertices on supporting branches or $v$ are reachable. The set of vertices visited during both traversals is exactly the set $C$ of candidates for the lowest common ancestor that we needed to find at this step.
\end{enumerate}

It is not hard to show that the algorithm indeed finds all such candidates $C$: it suffices to take any $w\in C$ and construct the paths $v'\rsqns w$ and $w\rsqns v''$ (or $w\rsqns v$): Algorithm~\ralg{step3} considers every edge of these paths in both traversals.

So, we have constructed an algorithm that finds the set of candidates $C$. Let us state the final theorem for this step:

\begin{stheorem}{candidatessimple}
Algorithm~\ralg{step3} finds the set of all non-supporting candidates for the lowest common ancestor such that there is a path $v'\rsq v$ containing this candidate. The running time of the algorithm is $O(\delta^{out}_G m_c)$.
\end{stheorem}
\begin{proof}
We proved the correctness of the algorithm earlier.
Let us deal with the running time.
In each connected component, the traversal takes $O(m_c)$, since we simply traverse the graph, with no need for prioritization. Since there are $O(\delta^{out}_G)$ components in total, the overall running time is $O(\delta^{out}_G m_c)$.
\end{proof}

\subsection{Speeding Up Step~3 in Practice}\label{sec:step3m}

We have constructed an algorithm for step~3 with a running time bound of $O(\delta^{out}_G m_c)$. The factor $\delta^{out}_G$ can be rather large, and from a practical standpoint the question arises: do we really need to traverse all non-supporting connected components containing vertices from $\sC(v')$? Can we enter only those from which we are guaranteed to reach the required $v''$?

Recall how Algorithm~\ralg{step3} works: first it traverses all non-supporting components from $v'$, reaches the supporting branches visited earlier by Algorithm~\ralg{step2}, collects the set of good reached supporting vertices $R$, and launches a traversal from $R$ in the reverse direction. The need to visit a large number of superfluous non-supporting connected components arises only during the first traversal. If we construct the set $R$ in a more efficient way, we can obtain a speedup in practice.

\sfigure{buildr}{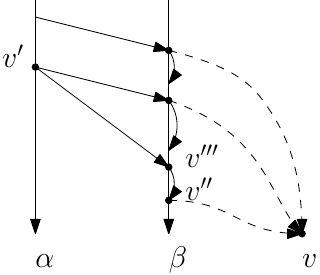}{Edges in the graph $\wG$ connecting $v'$ with vertices on the supporting branch $\beta$. $v''$ is the vertex on $\beta$ visited by the traversal ball from $v$. To enumerate all such edges starting from $v''$, it suffices to alternate transitions along augmentations of the second and third kinds along $\beta$ as long as there is an augmentation of the first kind from $v'$.}

In fact, everything we need to construct $R$ is already present in the graph $\wG$. Let us look carefully (see Fig.~\rfig{buildr}) at what the supporting branches $\alpha$ with $v'$ and $\beta$, by which Algorithm~\ralg{step2} reached $\alpha$, look like in the graph $\wG$. Algorithm~\ralg{step2} visits the supporting branch $\beta$ at the vertex $v''$, and, by the construction at step~2, we know that $v'\in\wB(v'')$. This means that either there is an augmentation of the first kind $(v',v'')$ in $\wG$, and then $v'\rsqns v''$, or there is an augmentation of the first kind $(v',v''')$ together with an augmentation of the second kind $(v''',v'')$. In the first case, we need to add $v''$ to $R$; in the second, $v'''$. But in $G$ there may be other non-supporting paths from $v'$ to ancestors of $v''$ on the supporting branch $\beta$. Recall that when we constructed $\wG$, we added augmentations of the first kind to all such vertices, and hence all of them can be traversed by alternating a transition along an augmentation of the second kind with a transition to the parent (augmentations of the third kind), as long as an augmentation of the first kind connects $v'$ with the visited vertex.

We obtain a simple Algorithm~\ralg{step3}' for constructing $R$. One needs to go over all supporting branches reached by Algorithm~\ralg{step2} and walk up along augmentations of the second and third kinds as long as an augmentation from $v'$ is preserved. Also, the start vertex $v$ may need to be included in $R$---but only if the supporting branch $\alpha$ was already reached at step~1 of the algorithm (i.e., if $v'$ is present in $S(v)$). We spend $O(1)$ time per candidate from $R$, and hence the time spent by the modified Algorithm~\ralg{step3} on constructing $R$ is no worse than that of the original one.

To ensure that the total running time of Algorithm~\ralg{step3}' is no worse than that of the original, one more detail is required. We no longer have the list $visited_{down}$ of vertices that were reached from $v'$, so when we start the traversal from a vertex of $R$, we need to choose correctly which non-supporting connected components to enter. Fortunately, we already have the computed $m_\alpha(\cdot)$ for all vertices---and we need to visit only those vertices for which $m_\alpha(\cdot) = h(v')$ (we certainly cannot get into vertices with $m_\alpha(\cdot) > h(v')$ at all, and vertices with $m_\alpha(\cdot) < h(v')$ are of no interest to us, since we need to reach $v'$).

Combining the above, we obtain a more pleasant running time bound:
\begin{stheorem}{step3m}
Algorithm~\ralg{step3}' finds the candidates for the lowest common ancestor in time $O(R^* m_c)$, where $R^*$ is the number of non-supporting connected components connecting $v'$ with the supporting branches reached by Algorithm~\ralg{step2} at vertices with height at most that of the reach vertex.
\end{stheorem}

\subsection{Finding the Lowest Common Ancestor among the Candidates (Step~4)}

After step~3, we have the set of candidates $C$, every vertex of which is certainly an ancestor of $v$, but is not necessarily an ancestor of $u$. The task of step~4 is to check whether $C$ contains ancestors of $u$ and, if so, to find the ancestor with the maximum height.

\sfigure{candcheck}{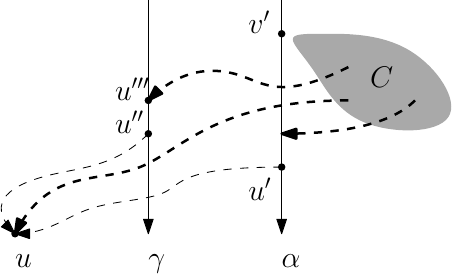}{Ancestors of $u$ from $C$ either reach $u$ via a non-supporting path, or reach the supporting branches visited by the traversal ball from $u$ at vertices with height smaller than that of the reach vertex.}

First, let us show that it suffices to check whether $C$ contains a common ancestor for the vertices from $reach[u]$. Indeed, suppose some vertex $w'\in C$ is reachable from $u$. Then, by the already familiar Lemma~\rlm{step2reach}, either $w'\rsqns u$, or there exist a supporting branch $\gamma$ and a vertex $u''\in\gamma$ such that $h(u'') > h(u''')$ and $reach[u, \gamma]=u''$, where $u'''$ is the first vertex of the intersection of $w'\rsq u$ and $\gamma$ (see Fig.~\rfig{candcheck}). Thus, it suffices for us to check reachability between $C$ and $reach[u]$ via non-supporting paths.

We use Algorithm~\ralg{step4}: this algorithm takes as input the set of candidate vertices $C$, traverses their descendants, stopping the traversal upon reaching a supporting branch, and finds the vertex of $C$ with the maximum height that is an ancestor of some of the reached supporting vertices whose height is at most $reach[u, \cdot]$ for the corresponding supporting branch. If no such vertices were found and $u$ was not visited either, the traversal guarantees that there is no common ancestor of the vertices $u$ and $v$ among $C$.

To understand the idea of the algorithm, consider Fig.~\rfig{checkidea}: for each vertex, we need to determine the candidate from $X$, $Y$, or $Z$ from which it is reachable, with the maximum height $h$. For this, it suffices to traverse the vertices in order of non-decreasing height and, for each one, take the maximum candidate among all its parents---this will be the desired candidate. 

\sfigure{checkidea}{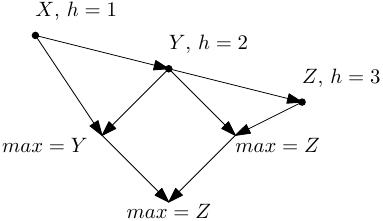}{Candidates $X, Y, Z$ and the maximum candidate computed for each vertex.}

\begin{salgorithm}{Step 4: finding the lowest common ancestor among the candidates $C$}{step4}
\State $C^* \gets \{x\ |\ \exists c\in C, c\rsqns x\}$
\State $candidate\_height[x], candidate\_result[x] \gets h(x), x\ \sforall\ x\in C^*$
\State $result\_height, result \gets 0, None$
\For{$x\in TopSort(G[C^*])$} \Comment\text{In order of non-decreasing height.} 
    \State $candidate\_result[x] \gets argmax_{y\in\sP(x)}(candidate\_height[y])$
    \State $candidate\_height[x] \gets max_{y\in\sP(x)}(candidate\_height[y])$
    \If{$x=u\text{\ \sor\ }(x\in S\text{\ \sand\ } \sigma(x)\in reach[u] \text{\ \sand\ } h(x) \le h(reach[u, \sigma(x)]))$}
        \State $result \gets argmax_{(result\_height,\ candidate\_height[x])} (result, candidate\_result[x])$
        \State $result\_height \gets max(result\_height,\ candidate\_height[x])$
    \EndIf
\EndFor
\State\Return $result$
\end{salgorithm}

To traverse the vertices in the required order, Algorithm~\ralg{step4} first performs a traversal in arbitrary order, starting from $C$ and stopping at supporting branches, and collects the set $C^*$ of visited vertices. Then, on the graph $G[C^*]$, a topological sort~\cite{Kahn} is performed. The resulting order allows the vertices to be traversed, visiting each one strictly after all of its parents from $C^*$. The first traversal, the topological sort, and the subsequent traversal in the required order take linear time. We obtain

\begin{stheorem}{step4}
Algorithm~\ralg{step4} finds the common ancestor of the vertices $u$ and $v$ from $C$ with the maximum height, or guarantees that there is none among $C$. The running time of the algorithm is $O(R^* m_c)$, where $R^*$ is the quantity from the statement of Theorem~\rthm{step3m}. In general form, the running time can be written as $O(\delta_G m_c)$.
\end{stheorem}
\begin{proof}
The correctness of the algorithm follows from the invariant: at the moment a vertex is visited (during the traversal in topological order), all of its parents reachable from $C$ have already been visited, and the candidate with the maximum height has been computed for each of them. Since the algorithm visits the vertices in order of non-decreasing height, the invariant holds.

The algorithm does not leave the non-supporting connected components. Constructing $C^*$, the topological sort, and the subsequent traversal take time linear in the number of edges. We obtain a running time of $O(R^* m_c)$ if we use the bound on the number of visited connected components, which depends on the particular pair $u$ and $v$. Or, in more general form, $O(\delta_G m_c)$.
\end{proof}

\subsection{Putting It All Together}

We have considered how the steps of our algorithm work separately; it remains only to put everything together. At the first step, we obtained the sets $S(u)$ and $S(v)$. At the second, we constructed $reach[u]$ and $reach[v]$, containing the reachable supporting branches and, for each, the vertex with the maximum height. If no common supporting branch was found, we are certain that no common ancestor of $u$ and $v$ exists. If there is one, we proceed to step~3, for which we compare the heights of the vertices at which the traversal balls reached the common supporting branch. If they arrived at the same vertex, steps~3 and~4 have to be repeated twice: for each of the two sides (see Fig.~\rfig{mirror}).

\sfigure{mirror}{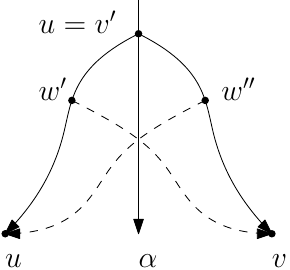}{If the traversal balls met on the supporting branch at the same vertex $v'$, then the lowest common ancestor may lie either on the path $v'\rsq v$ or on the path $v'\rsq u$.}

At step~3, we construct the set of candidates $C$, every vertex of which is certainly an ancestor of one of the vertices, and among which the lowest common ancestor is certainly present (if one exists). At the last step~4, we find the actual lowest common ancestor from $C$, or we can be certain that the vertex on the common supporting branch is the lowest common ancestor. Let us state the final theorem.

\begin{stheorem}{guaranteed}
The algorithm consisting of steps~1--4 finds the lowest common ancestor in time $O(R^* m_c + n_s\log n_s + m_s)$ (which, in more general form, can be written as $O(\delta_G m_c + n_s\log n_s + m_s)$), using $O(n_s)$ preprocessed memory per vertex.
\end{stheorem}
\begin{proof}
The correctness of the algorithm follows from the correctness of each step and the properties proved for these steps. Let us deal with the running time.
Step~1 can be performed in time $O(m_c)$ (or in $O(1)$ if $S(u)$ and $S(v)$ are precomputed).
By Theorem~\rthm{step2}, step~2 runs in time $O(n_s \log n_s + m_s)$.
By Theorems~\rthm{step3m} and~\rthm{step4}, steps~3 and~4 run in time $O(R^* m_c)$.
Adding everything together, we obtain $O(R^* m_c + n_s\log n_s + m_s)$.

The algorithm requires storing and precomputing the graph $\wG$, for which $m^*_\alpha(\cdot)$ is needed; also, at step~3, simply $m_\alpha(\cdot)$ is needed to direct the traversal. Storing $\wG$ requires $O(1)$ memory per supporting vertex for each other supporting branch, i.e., $O(n_s)$ memory per vertex in total. The same amount is needed for $m^*_\alpha(\cdot)$. For $m_\alpha(\cdot)$, $O(d)$ memory suffices (see Lemma~\rlm{mem1}).
\end{proof}

In Section~\rsec{memory}, we will show how this bound can be improved to $O(d)$ memory per vertex.

\section{Finding the Lowest Common Ancestor Without the Need to Traverse Edges from Parents to Children}\label{sec:noback}

The algorithm from Section~\rsec{guaranteed} has a significant drawback: it traverses vertices in order from parents to children. Fact~\rfact{fixedparent} guarantees that the set of parents of a vertex is fixed, but the set of children is not, and it changes as new vertices are added to the graph. Maintaining edge lists dynamically may be more costly; moreover, it simply requires twice as much memory for the edges.

Let us look more closely at the steps of the algorithm: steps~1 and~2 require traversal in order of non-increasing height; for step~3, we proposed an optimization in Subsection~\rsec{step3m}, which not only speeds up the process of constructing the candidate set $C$, but also makes it require exclusively transitions from vertices to parents. Step~4 remains. In step~4, we need to check which of the candidates in $C$ are ancestors of $u$, and for this we look at which of them have descendants on the supporting branches with height not greater than that reached by the traversal from~$u$ in the second step.

Imagine that for each non-supporting vertex $w$ we have computed $r_\alpha(w) = min(h(\{x|x\in\alpha, w\in \sAns(x)\}))$---the minimum height of a vertex on the supporting branch $\alpha$ for which $w$ is an ancestor. Then step~4 becomes trivial: we only need to iterate over the vertices from $C$ and, for each, compare $r_\alpha(\cdot)$ with the heights of the vertices reached by step~2 from $u$ over all supporting branches. If for some vertex $r_\alpha(\cdot)$ is not greater than $h(reach[u,\alpha])$, then this vertex is an ancestor of $u$.

\sfigure{reach}{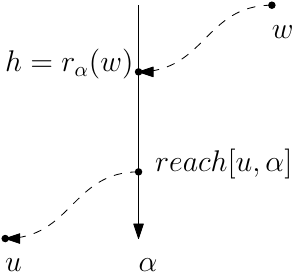}{Using $r_\alpha(w)$ to check whether $w$ is an ancestor of $u$.}

\begin{slemma}{noback}
The running time of the variant of step~4 that uses the precomputed $r_\alpha(\cdot)$ is $O(|C|d) = O(R^*m_cd)= O(\delta_Gm_cd)$.
\end{slemma}
\begin{proof}
For each non-supporting vertex from $C$, we need to compare its $r$ with the heights reached over all supporting branches, respectively. But only $O(d)$ supporting branches are reachable from a non-supporting vertex (both in a traversal from parents to children and from children to parents), and for the remaining ones $r_\alpha(\cdot)$ is not defined. This gives the statement of the lemma.
\end{proof}

It remains to show how $r_\alpha(v)$ can be constructed and maintained in reasonable time. We will show that this can be done in $O(d)$ amortized and $O(n_c)$ in the worst case.

\begin{stheorem}{addralpha}
Maintaining $r_\alpha(v)$ for all vertices and all supporting branches when a new vertex is added takes $O(d)$ amortized and $O(n_c)$ in the worst case.
\end{stheorem}
\begin{proof}
Suppose a new vertex $u$ is added to the graph. If it is non-supporting, then no values of $r_\alpha(\cdot)$ change. If $u$ is supporting, then perhaps for some vertices from $\sAns(u)$ we need to update $r_\alpha(\cdot)$. Note that $u$ currently has the maximum height among all vertices on $\alpha$, and therefore, if for some vertex $v$ the value $r_\alpha(v)$ is already defined, then $r_\alpha(v)$ does not need to be updated---the new value is certainly greater. Moreover, $r_\alpha(\cdot)$ does not need to be updated for all of $\sAns(v)$ either:
\begin{lemma}
If for some vertex $v$ the value $r_\alpha(v)$ is defined, then it is also defined for all $w\in \sAns(v)$.
\end{lemma}
\begin{proof}
By definition, there exists $u'\in\alpha$ such that $v\in\sAns(u')$. But then $w\in\sAns(u')$, and hence $r_\alpha(w) \le h(u') < h(u)$.
\end{proof}

Thus, when adding $u$, it suffices to traverse (via edges from child to parent) those non-supporting vertices for which $r_\alpha(\cdot)$ has not yet been defined. Moreover, Fact~\rfact{fixedparent} guarantees that visiting each vertex requires $O(1)$ time. It follows that adding $u$ requires time $O(|\{v|v\in\sAns(u), r_\alpha(v) = undefined\}|)$. We immediately obtain the second part of the statement---the time for maintaining $r_\alpha(\cdot)$ when adding a vertex takes $O(n_c)$ in the worst case.

It is easy to see (see Fig.~\rfig{reachupdate}) that if we sum the running time expressions over all $u\in\alpha$ over the entire running time of the algorithm, we simply obtain the number of non-supporting vertices (in fact, those from which $\alpha$ is reachable, but this is irrelevant right now). That is, the time for maintaining $r_\alpha(\cdot)$ over the entire running time of the algorithm is $O(n)$, but this means exactly that the amortized complexity of each individual operation is $O(1)$, if we speak only about a single supporting branch $\alpha$.

\sfigure{reachupdate}{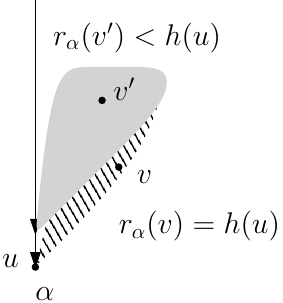}{Updating $r_\alpha(v)$ when a new vertex $u$ is added to the supporting branch $\alpha$.}

Since for each non-supporting vertex we need to maintain $O(d)$ values of $r$, i.e., over the entire running time of the algorithm it can be visited $O(d)$ times to update the corresponding $r$'s, we obtain an amortized complexity of $O(d)$.

(For greater rigor, one can use the amortized analysis method~\cite{Cormen} and say that adding each non-supporting vertex contributes $O(d)$ of potential, which is then spent on visits when updating $r$.)
\end{proof}

Combining the result of this section with Theorem~\rthm{guaranteed}, we obtain the theorem:
\begin{stheorem}{guaranteednoback}
Finding the lowest common ancestor can be performed in time $O(R^* m_cd + n_s\log n_s + m_s)$ (in a more general form: $O(\delta_G m_c d + n_s\log n_s + m_s)$), using $O(s)$ preprocessed memory per vertex and making transitions only from children to parents. The time required for maintaining $r$ when a new vertex is added to the graph is $O(d)$ amortized and $O(n_c)$ in the worst case. The amount of memory that must be dynamically updated in vertices is $O(d)$ per vertex.
\end{stheorem}

It may seem that we have not gained anything: the need to update vertices dynamically remains as before. But $r_\alpha$ has an advantage over $\sC$: the required amount of memory is bounded by $O(d)$, so it can be allocated immediately and the data can be written into already allocated slots.

\section{Memory Optimization}\lsec{memory}

So far, our graph $\wG$ requires $O(s)$ memory per supporting vertex. In this section, we will show how to reduce the consumption to $O(d)$ per vertex, without losing much in running time.

First, let us show that the in-degree of a supporting vertex in $\wG$ is bounded:

\begin{slemma}{memdegree}
For each supporting vertex, the number of augmentations of the first kind entering it in $\wG$ is bounded by $O(d)$.
\end{slemma}
\begin{proof}
Recall that augmentations of the first kind are edges connecting different supporting branches in $\wG$, and they correspond to non-supporting paths in $G$. Each supporting vertex has $O(1)$ parents (Fact~\rfact{smallparent}), and each non-supporting parent is reachable from $O(d)$ supporting branches (by the definition of $d$). It follows that the number of incoming edges connecting other supporting branches to a supporting vertex is bounded by $O(d)$.
\end{proof}

The graph $\wG$ contains augmentations of three kinds: between supporting branches, along a single supporting branch, and to parents on a supporting branch. We do not touch the augmentations between branches; they are essential, and they require $O(d)$ memory. Storing a supporting parent requires $O(1)$ memory. The augmentations along a supporting branch currently require $O(s)$ memory per vertex (since a separate transition is needed for each other supporting branch). We will try to store them differently.

Note that the augmentations along a single supporting branch are largely identical: they lead to the same vertex until a vertex with an augmentation from the corresponding other branch is encountered. If we simply store the transition not for every vertex but only in some of the vertices, we already reduce the memory consumption noticeably. That is, by paying $O(k)$ for moving along the supporting branch via parents until an augmentation is encountered, we reduce the memory by a factor of $k$, i.e., to $O(s/k)$ per vertex. If $d\approx \sqrt{s}$, we obtain $O(d)$ time and $O(d)$ memory per vertex.

But we also need to bound the memory consumption in the case where $d \ll \sqrt{s}$. Note that we have not yet used the fact that the number of augmentations between supporting branches (i.e., augmentations of the first kind) is rather small. Since there are $O(d)$ of them per vertex, for $b$ consecutive supporting vertices the number of incoming augmentations of the first kind is $O(db)$. Let us split the task of finding the augmentation we need into two:
\begin{enumerate}
  \item Finding the nearest augmentation inside a block of $b$ consecutive vertices.
  \item Finding the nearest block containing the required augmentation.
\end{enumerate}

Let us first deal with the search inside a block. Suppose we are working with the supporting branch $\sigma$. Then, for each supporting branch $\alpha\in\sS - \sigma$, write out into a list the augmentations from $\alpha$ to $\sigma$ in order of the height of the corresponding vertex on $\sigma$. In the block, for each augmentation, we store both endpoints of the edge---this way, we will be able to determine which vertex on $\sigma$ the augmentation corresponds to (and to find out the height of this vertex). We write into the block the lists for all supporting branches from $\sS - \sigma$ and add to them an index table that allows us to obtain the boundaries of the required list in time $O(1)$ (see Fig.~\rfig{memblock}). Then, to find the nearest augmentation from the corresponding supporting branch, we only need to find the position in the list corresponding to the height of the current vertex---the previous element of the list (if it exists) will be the required augmentation. The time for searching in a sorted array is $O(\log b)$~\cite{Knuth}.

\sfigure{memblock}{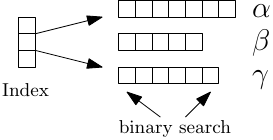}{Organization of the storage of augmentations inside a block.}

Now let us understand what to do when the current block contains no corresponding augmentation. We would like to jump directly to the required block, and it may be rather far from the current one. Let us add special skip-blocks between blocks, in which, for all $O(s)$ supporting branches, we store a reference to the block with the nearest augmentation to it (see Fig.~\rfig{memskip}).

\sfigure{memskip}{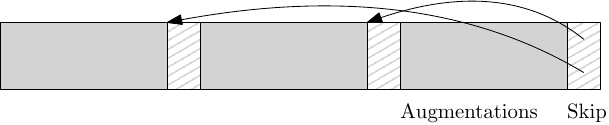}{Transitions to the required block.}

Let us now compute how much memory is required. The size of a block with augmentations is $O(db + s)$ (the augmentations themselves and the index table), and additionally $O(s)$ per block for the skip-block. With $b\approx s$, we obtain $O(d)$ memory per vertex and $O(\log s)$ time to find the nearest augmentation.

By Lemma~\rlm{mem1}, when a new supporting vertex is added to the graph, we only need to add $O(d)$ elements to the lists in the last block, which takes time $O(d)$. Maintaining $m_\alpha$ and $m^*_\alpha$ also takes time $O(d)$. By Theorem~\rthm{guaranteednoback}, maintaining $r_\alpha$ also requires $O(d)$ amortized time.

Combining the result of this section with Theorems~\rthm{guaranteed} and~\rthm{guaranteednoback}, we obtain:

\begin{stheorem}{guaranteedod}
Finding the lowest common ancestor can be performed in time $O(\delta_G m_c + n_s\log^2 n_s + m_s\log n_s)$ with $O(d)$ preprocessed memory per vertex, while the time for adding a new vertex to the graph is $O(d)$. If only transitions from child to parent are allowed, then finding the lowest common ancestor can be performed in time $O(\delta_G m_c d + n_s\log^2 n_s + m_s \log n_s)$ with $O(d)$ preprocessed memory per vertex, while the time for adding a new vertex to the graph is $O(d)$ amortized and $O(n_c)$ in the worst case, and additionally $O(d)$ of dynamically updated memory per vertex must be maintained.
\end{stheorem}

\end{document}